\documentstyle[12pt]{article}
\newcommand{\ber}{\begin{eqnarray}}
\newcommand{\eer}{\end{eqnarray}}
\newcommand{\bea}{\begin{equation}}
\newcommand{\eea}{\end{equation}}
\begin{document}
\title{Fermion to Boson Mapping}
\author{ Asis Basu and Jayprokas Chakrabarti \\
\small
 Theory Group\\
\small
Indian Association for the Cultivation of Science\\
\small
 Calcutta-32    INDIA.\\
 \small}
\date{}
 \maketitle

\begin{abstract}
The enveloping algebra,$D_{n}$,of fermions is extended on the
lattice to include the discrete space invariance.This extended
algebra,denoted  X, has the space symmetry as a factor :
$X/D_{n}$ = space group.
\end{abstract}
\maketitle
\section{Introduction}
In a recent couple of papers [1,2] we expressed the dynamics of
fermions moving on the linear array of molecules in terms of
bosonic bonds between  the fermions. Boson realisations of Lie
algebras have seen wide applications in many areas[3].Earlier it
used to be in continuum theories, but now the goal is to go from
continuum to lattice models[4] to solve them on the computer,for
instance, in the quark-glue systems.The enveloping algebra for the
case of spinless fermions on m-lattice sites is $A_{2m-1}$ if it
conserves the particle number.In the number non-conserving case it
is $D_{2m}$.On the lattice both these have to extend to include
the discrete space symmetry.
\newline
 In this paper we extend our results [1,2] to two cases : i) when the simple translation
by one lattice site is not the symmetry of the Hamiltonian and
ii) in more than one dimensions as well as in arbitrary large
dimensional systems.Our efforts, much like in [1,2] , is to
transform the simple Hamiltonian of fermions to those of bond
bosons.We indicate briefly how to write the fermionic interactions
in terms of these bosons.
\newline
This mapping from the problem of fermions to the problem of bosons
is motivated by the following reasons:
\smallskip
a) In some situations, such as in the harmonic oscillator
potential, the bosonic co-ordinates are physically more
relevant[5].Some of the n-point fermion functions have poles that
are bosonic.In these cases the theory below some threshold
temperature is dominated by these bosons[3].
\smallskip
b) There are other cases where the fermion problem may be mapped
to some equivalent boson problem.This mapping could simplify the
problem somewhat,for example,reducing the quartic fermion terms to
quadratic in  bosons etc.There have been wide uses of these types
of boson-mapping in a variety of problems, both in condensed
matter theory [6] as well as in other many-body systems[3].
\newline
There have been instances where the distinction between (a) and
(b) disappears. The boson mapping of fermion theories generate
approximate co-ordinates that are physically relevant. One such
example is the Holstein-Primakoff mapping [7] of spin systems in
terms of bosons.These bosons are the approximate physical degrees
of freedom at least for high-spin Heisenberg interactions.In this
paper we explore fermionic systems mapped to bosonic bonds.
\newline
The mapping of fermions to bosons is a subject of interest on its
own. As early as 1950 ,Tomonaga transformed the usual quartic term
describing the interaction of electrons to  terms in density
fluctuations that were treated as bosons. The problem of
interacting electron gas has been mapped to these bosons in one
dimension [6] by Tomonaga , Luttinger , Mattis and others.In
recent years the properties of fermions in low- dimensions have
been a subject of great interest[8,9] .
\newline
The Fourier transform of the 2-point fermion correlation function
gives the momentum distribution n(k).This quantity has a finite
discontinuity at the fermi surface for fermi-liquids.This
discontinuity is called the quasiparticle renormalisation factor
$z_{k}$ . In non-interacting theories $z_{k}=1$ . As the
interactions are turned on , $0<z_{k}<1$. The 2-point fermion
function :
\begin{equation}
G(x-x',t-t')= <0|T\psi(x,t)\psi^{\dag}(x',t')|0>
\end{equation}
 has the Fourier transform :
\begin{equation}
\widetilde{G}(k,\omega)=\frac{i}{\omega-\epsilon_{k}-\Sigma(k,\omega)}
\end{equation}
where the chemical potential is absorbed in defining
$\epsilon_{k}$ so that $\epsilon_{k}=0$ for k on the fermi
surface.The $\Sigma(k,\omega)$ , called the self-energy,
quantifies the effects of the interactions and also contains the
prescription necessary to shift the pole off the real axis for
$\omega$ . In case of fermi-liquids,
\begin{equation}
z_{k}=(1-\frac{\partial\Sigma}{\partial\omega})^{-1}
\end{equation}
 gives the discontinuity of n(k) on the Fermi surface.
It is known that fermion systems in low-dimensions, often behave
otherwise.At the fermi-surface in these cases there is no
discontinuity in the momentum distribution function.Instead there
is a cusp whose form is determined from an exponent which depends
on the interactions.This governs the power-law fall offs of the
n-point functions as the space-time distances increase[10].
\newline
 For the mapping of fermions to bosonic bonds we considered in [1,2] an
one dimensional array of atoms/molecules on which the fermions
move. This chain is assumed to have 2N sites.[It is not
necessary to consider the site number to be even ; it could
equally well be odd.] Consider the bond variables :
\begin{equation}
e_{+l}=\sum c_{n}^{\dagger} c_{n+l}^{\dagger}
\end{equation}
where $ c_{m}^{\dagger}$ and $c_{m}$ creates and destroys the
fermion at site m. This m takes values from 1 to 2N. Clearly,

\begin{equation}
\{c_{m},c_{n}^{\dagger}\}=\delta_{mn}
\end{equation}
\begin{equation}
\{c_{m},c_{n}\}=\{c_{m}^{\dagger},c_{n}^{\dagger}\}=0
\end{equation}
Note that in (4) we have paired fermions separated by l-sites and
superposed. The reason for this superposition is to generate the
object $e_{+l}$ [1] which has translational invariance on the
one-dimensional lattice. Our assumption here has been that the
underlying dynamics i.e. the Hamiltonian ,is invariant under
translation  by one site.
\smallskip
There are two points to keep in mind at this stage :
\newline
(1) that the object $e_{+l}$ need not be invariant of the
space group of the lattice.It is sufficient if it is an
irreducible representation of that group.
\newline
(2) Even though the lattice has periodicities of translation by a
site, the dynamics - the Hamiltonian - may not have that
symmetry.In these cases  it is convenient  to define the bond
variables  with the underlying invariance.
\newline
Amongst the Vinyl Polymers Polyacetylene (PA) is a member made of
$(CH)_{X}$. Each CH gives one $\pi$- electron ; therefore the band
is half-filled . The lattice-fermion interaction[11,12] makes this
partially filled band unstable with respect to a lattice
distortion at $2k_{F}$. ; $k_{F}$ being the Fermi vector.
Clearly, the Hamiltonian for this system is not invariant under
translation by a single lattice site.Interestingly, the solutions
of the Dirac Hamiltonian on the lattice are the long wavelength
modes of the Hamiltonian for PA of Su-Schriffer-Heeger
(SSH)[13,14]. We take up this case,define the  boson-bond mapping
and solve the SSH-Hamiltonian in terms of these bosons.
\newline
Coming  back now to point (1) that the bosonic bonds need not be
invariant with respect to  the underlying space-group. To
generate the ireducible representation we "distorted"[1,2] the
superposed  fermion pairs(l-sites apart) by defining the
variables
\begin{equation}
e_{+lk}=\sum exp^{ikn}.c_{n}^{\dagger}c_{n+l}^{\dagger}
\end{equation}
The quantities k are obtained using periodic boundary condition,
$exp^{ik.2N}=1$. Note we could have used the antiperiodic boundary
 condition as well.  We  showed in [1] and [2] that the quatities  $e_{+lk}$
along  with their conjugates $e_{-lk}=(e_{+lk})^{\dagger}$
satisfy an approximate bosonic commutator relationship, when
properly normalised :
\begin{equation}
[\frac{e_{+lk}}{\sqrt{2N}},
\frac{e_{-lk}}{\sqrt{2N}}]=\delta_{ll^{\prime}}\delta_{kk^{\prime}}
\end{equation}
in the filling region where the fermion states are all nearly
occupied . In this state $e_{+lk}$ are the destruction operators of
the bosonic bonds; $e_{-lk}$ create them . The Hamiltonian of
fermions is transformed to the Hamiltonian of the bosonic bond
co-ordinates $e_{\pm lk}$ , much like what Kronig[15] did in terms
of the bosonic density fluctuations.The transformed Hamiltonian of
the bosonic-bond co-ordinates is then solved.
\smallskip
We discussed these solutions in a few cases . The consistency of
this approach is brought out by comparing the eigenstates of of
bosonic bonds with respect to the solutions for the known
fermionic problem.These comparisons convinced us of the
reliability of the approximations.The procedure is extended here
for the following cases:
\newline
a) The dynamics due to variety of reasons may not have the
symmetry  of  the underlying  lattice, i.e.periodicity w.r.t.
translation by a single-site.We  take up the well-understood case
of  PA in order to  be  able to compare  our bond solutions to
the known  solutions in terms  of the  fermion.
\newline
b) to extend the considerations to higher dimensions. We define
the bosonic bonds in higher dimensions and solve a case of
fermionic Hamiltonian in terms of bosonic bonds in (2+1) D.
\newline
c) We discuss briefly the transformation of quartic fermionic
interactions to quadratic terms in bosonic  bonds.

\section{Boson Mapping}
Denote the algebra of the fermion bilinears by X.On the
one-dimensional lattice of N-sites,these bilinears are of the type
$c_{i}^{\dag}c_{j}^{\dag}$, $c_{i}^{\dag}c_{j}$ and $c_{i}c_{j}$
(where  i,j take values from 1 to N).These bilinears go into one
another under the action of the discrete space invariance.For the
N-point lattice, for the spinless case,the algebra of the fermion
bilinears is isomorphic to $D_{N}$. If we include spin it is $D_{2N}$.
 Clearly $D_{2N} (orD_{N})$ is
an invariant subalgebra of X because the action of the space group
does not take the generators out of $D_{2N} (orD_{N})$. Hence
$X/D_{N}$= space group, for the spinless case;$X/D_{2N}$= space
group, for the fermions with spin.The induced representations are
the ones that become relevant as illustrated here.

 Around 1958 Natta, Mazzanti and Corradini converted the
monomer acetylene  $C_{2}H_{2}$  into polymer PA. Because of the
fermion-lattice interactions an energy  gap develops  at $\pm
k_{F}$ and PA is a semiconductor.Sometimes the resulting
distortion is  called dimerisation that  is  the alternate short
and long  bonds.In our study we  keep in  mind these lattice
distortions. The  bond bosons we construct have  to be
irreducible representations of the space group of this distorted
lattice.The monomers CH have six degrees of  freedom.Only  one of
these , the translation along the chain direction is important in
dimerisation.It is  the one kept  in the Hamiltonian.The
fermion-lattice coupling in the tight-binding approximation,due to
Ovchinnikov[11,12] , has a more general form than the Fr\"{o}hlich
Hamiltonian and reduces to it in the $k\rightarrow 0$ limit.This
Hamiltonian was discussed by Su-Schrieffer-Heegger[13] to explore
the soliton states in PA.
\newline
This Hamiltonian is a simple example of the case that it does not
commute with translation by a single site :
\begin{equation}
H = -t_{0} \sum (c_{n\sigma}^{\dagger}c_{(n+1)\sigma}) +
2\alpha.u \sum
[(-1)^{n}.c_{n\sigma}^{\dagger}c_{(n+1)\sigma}]+h.c.
\end{equation}
We construct the bond bosons to reflect the underlying lattice
space group.Divide the lattice into odd(A) and the even(B) lattice
and define the bond bosons as follows :
\begin{eqnarray}
e_{+lk}^{A}=\sum_{n=odd}
c_{n}^{\dag}c_{n+l}^{\dag}e^{ikn}\\
e_{+lk}^{A}=\sum_{n=even}
c_{n}^{\dag}c_{n+l}^{\dag}e^{ik\frac{n}{2}}
\end{eqnarray}
The symbol $c_{n}^{\dagger}(c_{n})$ denotes the
creation(destruction) operator for the up-spin fermion at site
n.The corresponding operators for spin down fermions at the same
site are denoted by $d_{n}^{\dagger}(d_{n})$ respectively. Similar
bond- bosons with $c\rightarrow d$ (i.e. spin-up replaced by  spin
down fermions.) are:
\begin{eqnarray}
\overline{e_{+lk}^{A}}=\sum_{n=odd} d_{n}^{\dag}d_{n+l}^{\dag}e^{ikn}\\
\overline{e_{+lk}^{B}}=\sum_{n=even}
d_{n}^{\dag}d_{n+l}^{\dag}e^{ik\frac{n}{2}}
\end{eqnarray}
Note that in writing these operators we have used the idea that
under translation by 2-lattice sites the $e_{lk}^{A(B)}$ goes
over to $e_{lk}^{A(B)}$ upto an overall phase. This ensures that
these are irreducible representations of the underlying space
group.

The bond bosons listed above do not exhaust all the
possibilities.The bonds between spin-up fermions and spin-down
fermions are bosonic as well. These are listed hereunder:
\begin{eqnarray}
e_{+lk}^{\prime A} =\sum_{n=odd} c_{n}^{\dagger}d_{n+l}^{\dagger}e^{ikn}\\
e_{+lk}^{\prime B}=\sum_{n=even} c_{n}^{\dagger}d_{n+l}^{\dagger}e^{ik\frac{n}{2}}\\
\overline{e}_{+lk}^{\prime A}=\sum_{n=odd}d_{n}^{\dag}c_{n+l}^{\dag}e^{ikn}\\
\overline{e}_{+lk}^{\prime
B}=\sum_{n=even}d_{n}^{\dag}c_{n+l}^{\dag}e^{ik\frac{n}{2}}
\end{eqnarray}

We note that the SSH Hamiltonian is spin independent. That means
that bosonic bonds could be between fermions of the same spin or
between fermions of opposite spin. We shall discuss the physical
relevance of these possibilities later in this section.

It is convenient to deal with the linear combinations defined as :
\begin{eqnarray}
E_{\pm lk}^{A(B)} (\pm)= e_{\pm lk}^{A(B)}  \pm \overline{e}_{\pm
lk}^{A(B)} \\
D_{\pm lk}^{A(B)} (\pm)= e_{\pm lk}^{\prime A(B)} \pm
\overline{e}_{\pm lk}^{\prime A(B)}
\end{eqnarray}
The subscript minus denotes the object that is hermitian
conjugate of the one with subscript plus .
\newline
To calculate the dynamics of the bond bosons $E_{\pm lk}^{A(B)}
(\pm)$ and $D_{\pm lk}^{A(B)} (\pm)$, we calculate  the
commutators with the Hamiltonian (9). We get:
\begin{equation}
[H,E_{+lk}^{A}] =
E_{+(l+1)k}^{A}.t_{0,l+1}+E_{+(l-1)k}^{A}.t_{0,l} +
E_{+(l+1)k}^{B}.t_{0,+}e^{ik}+E_{+(l-1)k}^{B}.t_{0,-}
\end{equation}
\begin{equation}
[H,E_{+lk}^{B}] =
E_{+(l+1)k}^{B}.t_{0,l}+E_{+(l-1)k}^{B}.t_{0,l+1} +
E_{+(l+1)k}^{A}.t_{0,-}+E_{+(l-1)k}^{A}.t_{0,+}e^{-ik}
\end{equation}
\begin{equation}
[H,D_{+lk}^{A}] =
D_{+(l+1)k}^{A}.t_{0,l+1}+D_{+(l-1)k}^{A}.t_{0,l} +
D_{+(l+1)k}^{B}.t_{0,+}e^{ik}+D_{+(l-1)k}^{B}.t_{0,-}
\end{equation}
\begin{equation}
[H,D_{+lk}^{B}]  =
D_{+(l+1)k}^{B}.t_{0,l}+D_{+(l-1)k}^{B}.t_{0,l+1} +
D_{+(l+1)k}^{A}.t_{0,-}+D_{+(l-1)k}^{A}.t_{0,+}e^{-ik}
\end{equation}
where $t_{0,l}=t_{0}+(-1)^{l}.2\alpha u$ and $t_{0,\pm}=t_{0}\pm
2\alpha u$ The commutators with the conjugates $E_{-lk}^{A(B)}
(\pm)$ and $D_{-lk}^{A(B)} (\pm)$,may be obtained by  taking the
hermitian conjugates of the above four relations. We note that the
bond objects $E_{\pm lk}^{A(B)} (\pm)$ and $D_{\pm lk}^{A(B)}
(\pm)$ are bosonic.Using the bosonic commutator relations we can
transform the Hamiltonian (9) written in terms of fermions into an
equivalent Hamiltonian of bosonic bonds.

This equivalent bosonic bond Hamiltonian is :
\begin{eqnarray}
H_{B}(E)= \sum_{k}
[t_{0,l+1}(E_{+(l+1)k}^{A}E_{-lk}^{A}+E_{+lk}^{A}E_{-(l+1)k}^{A})
+t_{0,l}(E_{+(l+1)k}^{B}E_{-lk}^{B}+E_{+lk}^{B}E_{-(l+1)k}^{B})]
\nonumber \\
+[t_{0+}(e^{ik}E_{+(l+1)k}^{B}E_{-lk}^{A}+e^{-ik}E_{+lk}^{A}E_{-(l+1)k}^{B})
+t_{0-}(E_{+lk}^{B}E_{-(l+1)k}^{A}+E_{+(l+1)k}^{A}E_{-lk}^{B})]
\end{eqnarray}
We have in addition the terms of the D-sector . The Hamiltonian in
the D-co-ordinates is denoted by $H_{B}(D)$.The subscript B is to
emphasize that these are bosonic Hamiltonian.
 Now we want to diagonalize the complete bond bosonic Hamiltonian:
 $H_{B}=H_{B}(E)+H_{B}(D)$
 To diagonalize $H_{B}(E)$ we go over to q-space by Fourier
 transformation.The procedure below for$H_{B}(E)$ ; for $H_{B}(D)$
 it is similar.
 Define
 \begin{equation}
 E_{\pm lk}^{A(B)} = \frac{1}{\sqrt L} \sum E_{\pm qk}^{A(B)}.
 e^{\mp iql}
 \end{equation}
\newpage
 Carrying out this Fourier transform over$H_{B}(E)$ we get
\begin{eqnarray}
H_{B}(E)&=& \sum_{q,k} [ E_{+qk}^{A}.E_{-qk}^{A} +
E_{+qk}^{B}.E_{-qk}^{B}]2t_{0}\cos q +\nonumber \\
& & [E_{+qk}^{A}.E_{-(q-\pi)k}^{A} -
E_{+qk}^{B}.E_{-(q-\pi)k}^{B}]4\alpha ui\sin q +\nonumber \\
& & E_{+qk}^{B}.E_{-qk}^{A}[e^{\frac{ik}{2}}(2t_{0}\cos{(k/2-q)}
+4\alpha ui\sin{(k/2-q)})] \nonumber \\
& & +E_{+qk}^{A}.E_{-qk}^{B}[e^{-\frac{ik}{2}}(2t_{0}\cos{(k/2-q)}
-4\alpha ui\sin{(k/2-q)})]
\end{eqnarray}
We now set  $z=e^{\frac{ik}{2}}(2t_{0} \cos(k/2-q) +i.4\alpha
u\sin(k/2-q)$ . Therefore the Hamiltonian $H_{B}(E)$ to
diagonalize becomes
\begin{eqnarray}
H_{B}(E)  =  \sum [ E_{+qk}^{A}.E_{-qk}^{A} +
E_{+qk}^{B}.E_{-qk}^{B}]y +  [E_{+qk}^{A}.E_{-(q-\pi)k}^{A} -
E_{+qk}^{B}.E_{-(q-\pi)k}^{B}]iX \nonumber \\ +
E_{+qk}^{B}.E_{-qk}^{A}z +E_{+qk}^{A}.E_{-qk}^{B}\overline{z}
\end{eqnarray}
where $y=2t_{0}\cos q$ , $X= 4\alpha u \sin q$, $\overline{z}$ is
complex conjugate of z. In matrix form :
\begin{equation}
\left(\begin{array}{cccc} y & -iX & \overline{z} & 0 \\iX & -y & 0
& -\overline{z} \\z & 0 & y & iX \\0 & -z & -iX & -y
\end{array}\right)
\end{equation}
If we call the above matrix A , its eigenvalues $\lambda$ are of
interest.$A-\lambda.I=0$ gives us the following  four values for
$\lambda$: \bea \lambda=\pm \sqrt{(2t_{0}\cos
q)^{2}+(4\alpha.u\sin q)^{2}}\mp \sqrt{(2t_{0}\cos
(k/2-q))^{2}+(4\alpha.u\sin(k/2-q))^{2}} \eea Let us now compare
the above boson-bond spectrum with that of single fermions[13].The
lattice-fermion interaction leads to a band gap , parametrized by
$\triangle_{K}$ , given by
\begin{equation}
 \triangle_{K}= 4\alpha.u.\sin K
\end{equation}
(Note that we have the lattice spacing a=1). The symbol K is to
distinguish it from k that we have used for the boson case.The
relation between k and K will be discussed shortly. The single
fermion energy spectrum is given by
 \bea
 E_{K}=\pm \sqrt{\epsilon_{K}^{2}+\triangle_{K}^{2}}
\eea
 where
\begin{equation}
\epsilon_{K}=2t_{0} \cos K
\end{equation}
  with a,the lattice spacing is set to 1 and $\triangle_{K}$ given by (30).The upper spectra,
 the + sign in (31), is  the conduction band ; the lower one the valence
band.Before we compare this single-fermion spectrum with that of
the bosons,eqn.(29), let us think in terms of a small PA chain
$(CH)_{6}$ -i.e. the lattice of 6-sites.The values of k are
obtained from $e^{3ik}=1$. The values of q(eqn(25)) are obtained
from the periodic boundary condition on the boson lattice[1].Since
for the lattice of 6-points there are 3 independent values for
l,the condition on q is:
\begin{equation}
e^{3iq}=1
\end{equation}
 For the 6-point $(CH)_{6}$ chain, the K values are
0,$\frac{2\pi}{3}$ and $\frac{4\pi}{3}$.Thus K,k and q range over
the same values. Looking now at the bosonic bond spectrum (29), we
notice that depending on the choice of signs in (29),the bonds
could be between two valence  band fermions, or between two
conduction band fermions or between a fermion in the valence band
with another in the conduction band.The bond spectrum is in one to
one correspondence with the single fermion spectrum. Hence the two
descriptions are equivalent.This shows that the program that we
have laid out for the construction of the bonds is consistent. The
Hamiltonian $H_{B}(E)+H_{B}(D)$ is spin-independent.Therefore, the
D-sector(i.e. bonds between the up and the down spins) is
identical to the E-sector(i.e. bonds between spins of the same
type)Diagonalisation of the D-sector gives the same result as
(29).

\section{In Higher Dimensions}
We extend in this section the concept of bosonic bonds to higher
dimensions. In the 1-dim. case  we transformed the Hamiltonian of
fermions to the Hamiltonian of bosonic bonds. In the previous
section we discussed the invariant subalgebra  $D_{n}$ of X and
the factor - the space invariance.The induced representations are
important.We solved the Hamiltonian of bosonic bonds and compared
with known solutions of the fermions to check for the consistency
of our approach.The bosonic bonds satisfy the bosonic commutator
rules; much like the density fluctuations.We want to do the same
for higher dimensions.The construction of the bonds involved
pairing two fermions separated l-sites apart and superposing with
similar other fermion pairs such that the resulting bond variable
is an irreducible representation of the underlying space group.
\newline
 To extend to two dimension consider the
simple  square lattice on which the fermions move. We define the
bond boson operator as :
\begin{equation}
e_{+lm}(k_{x},k_{y})=\Sigma
c_{xy}^{\dag}c_{x+l,y+m}^{\dag}exp^{i(xk_{x}+yk_{y})}
\end{equation}
along with its conjugate $e_{-lm}(k_{x},k_{y})$
\newline
The quantities $k_{x}$ and $k_{y}$ are determined by the p.b.c.
 The algebra of
$e_{+lm}(k_{x},k_{y})$ closes [1] with another type of operators
$h_{\pm ab}(k_{x},k_{y})$ defined as :
\begin{equation}
h_{+ab}(k_{x},k_{y})=\Sigma
c_{xy}^{\dag}c_{x+a,y+b}exp^{i(xk_{x}+yk_{y})}
\end{equation}
with $h_{-ab}(k_{x},k_{y})=(h_{ab}(k_{x},k_{y}))^{\dag}$ Using
our arguments as earlier [1,2] it is then straightforward to
establish that
\begin{equation}
[e_{+lm}(k_{x},k_{y}),e_{-lm}(k_{x},k_{y})]
=(2N)^{2}\delta_{ll'}\delta_{mm'}\delta_{k_{x}k_{x}'}
\delta_{k_{y}k_{y}'}
\end{equation}
near the filling region where all the fermion states are occupied.
Thus if we normalise as $e_{+lm}(k_{x},k_{y})\rightarrow
\frac{1}{2N}e_{+lm}(k_{x},k_{y})$ these normalised bond bosons
satisfy the standard bosonic commutators.
\newline
 Setting up the Dirac
Hamiltonian in two-dimension requires a two component wave
function $\psi_{1}(\overrightarrow{x})$at each site
$\overrightarrow{x}=(x_{1},x_{2})$ defined as:
\begin{equation}
\left(\begin{array}{c} \psi_{1}(\overrightarrow{x})\\
\psi_{2}(\overrightarrow{x})
\end{array}\right)
\end{equation}
The Hamiltonian with mass term
\begin{equation}
H=\sum_{\overrightarrow{x}}i\psi^{\dag}(\overrightarrow{x})(\alpha.\partial)\psi(\overrightarrow{x})+
\Delta
\psi^{\dag}(\overrightarrow{x})\beta\psi(\overrightarrow{x})
\end{equation}
for the choice : $\alpha_{1}=-\sigma_{2}$
;$\alpha_{2}=\sigma_{1}$; $\beta=\sigma_{3}$ can be written on the
lattice as :
\begin{eqnarray}
 H=\sum_{x,y}
c_{xy}^{\dag}([b_{x-1,y}-b_{x+1,y}]+i[b_{x,y+1}-b_{x,y-1}])\nonumber\\
+b_{xy}^{\dag}([c_{x+1,y}-c_{x-1,y}]+i[c_{x,y+1}-c_{x,y-1}])\nonumber\\
+\Delta(c_{xy}^{\dag}c_{xy}-b_{xy}^{\dag}b_{xy})
\end{eqnarray}
where we have identified $c_{xy}$  and $b_{xy}$ with
$\psi_{1}(\overrightarrow{x})$ and $\psi_{2}(\overrightarrow{x})$
 respectively.
 Let us construct the operators:
\begin{eqnarray}
e_{+lm}^{1}(k)=\sum
c_{xy}^{\dag}c_{x+l,y+m}^{\dag}exp^{i(k_{x}x+k_{y}y)} \\
e_{+lm}^{2}(k)=\sum
b_{xy}^{\dag}b_{x+l,y+m}^{\dag}exp^{i(k_{x}x+k_{y}y)} \\
\overline{e}_{+lm}^{1}(k)=\sum
c_{xy}^{\dag}b_{x+l,y+m}^{\dag}exp^{i(k_{x}x+k_{y}y)} \\
\overline{e}_{+lm}^{2}(k)=\sum
b_{xy}^{\dag}c_{x+l,y+m}^{\dag}exp^{i(k_{x}x+k_{y}y)}
\end{eqnarray}
Now , as earlier, build the linear combination
\begin{eqnarray}
E_{\pm lm}^{\pm 1}(k)=\frac{1}{\sqrt{2}}[e_{\pm lm}^{1}(k)\pm
e_{\pm lm}^{2}(k)]\\
E_{\pm lm}^{\pm 2}(k)=\frac{1}{\sqrt{2}}[\overline{e}_{\pm
lm}^{1}(k)\pm \overline{e}_{\pm lm}^{2}(k)]
\end{eqnarray}
The $E_{\pm lm}$ operators satisfy bosonic commutator rules.
\newline

Let us now calculate the commutators of these operators with the
the concerned Hamiltonian, so that we get the following
successive results :
\begin{eqnarray}
[H,E_{+lm}^{\pm 1}(k)] = E_{(l-1)m}^{\mp 2}(k)(1\mp exp^{-ik_{x}})
\mp E_{(l+1)m}^{\mp 2}(k)(exp^{ik_{x}}\mp 1) \nonumber \\
 + iE_{l(m-1)}^{\pm 2} (k)(1+ exp^{-ik_{y}})-iE_{l(m+1)}^{\pm 2}(k)(1+ exp^{ik_{y}})
-2m E_{+lm}^{\mp 1}(k)
\end{eqnarray}
and
\begin{eqnarray}
[H,E_{+lm}^{\pm 2}(k)] = E_{(l+1)m}^{\mp 1}(k)(1\mp exp^{ik_{x}})
\mp E_{(l-1)m}^{\mp 1}(k)(1\mp exp^{-ik_{x}})\nonumber \\
 + iE_{l(m-1)}^{\pm 1} (k)(1+ exp^{-ik_{y}})-iE_{l(m+1)}^{\pm
1}(k)(1- exp^{ik_{y}})
\end{eqnarray}
From the above two commutator results the Hamiltonian for the
bond bosons can be written as :
\begin{eqnarray}
 H_{B}= (1-e^{ikx})[E_{+(l+1)m}^{-1}E_{-lm}^{+2} - E_{+(l+1)m}^{+2}E_{-lm}^{-1}]\nonumber \\
+(1+e^{ikx})[E_{+(l+1)m}^{+1}E_{-lm}^{-2} - E_{+(l+1)m}^{-2}E_{-lm}^{+1}]\nonumber \\
+i(e^{iky}-1)[E_{+l(m+1)}^{+1}E_{-lm}^{+2} - E_{+l(m+1)}^{+2}E_{-lm}^{+1}]\nonumber \\
-i(e^{iky}+1)[E_{+l(m+1)}^{-2}E_{-lm}^{-1}+E_{+l(m+1)}^{-1}E_{-lm}^{-2}]\nonumber \\
-2m E_{+lm}^{-1}E_{-lm}^{+1}
+h.c.
\end{eqnarray}
  We go over to Fourier transformed space(s-p) and define:
\begin{equation}
E_{\pm l,m}^{\pm 1} = \frac{1}{LM}\sum M_{\pm sp}^{\pm}e^{\mp
isl}e^{\mp ipm}
\end{equation}
\begin{equation}
E_{\pm l,m}^{\pm 2} = \frac{1}{LM}\sum N_{\pm sp}^{\pm}e^{\mp
isl}e^{\mp ipm}
\end{equation}
In terms of these new variables the equivalent Hamiltonian for
this 2D case is:
\begin{eqnarray}
H_{B}= \sum_{sp}
N_{+sp}^{+}M_{-sp}^{-}[e^{is}-e^{-is}+e^{i(k_{x}-s)}-e^{-i(k_{x}-s)}]\nonumber \\
+M_{+sp}^{+}N_{-sp}^{-}[e^{-is}-e^{is}+e^{-i(k_{x}-s)}-e^{i(k_{x}-s)}]\nonumber \\
+iM_{+sp}^{+}N_{-sp}^{+}[e^{ip}-e^{-ip}+e^{i(k_{y}-p)}-e^{-i(k_{y}-p)}]\nonumber \\
+iN_{+sp}^{-}M_{-sp}^{-}[e^{ip}-e^{-ip}+e^{-i(k_{y}-p)}-e^{i(k_{y}-p)}]\nonumber \\
-2m [M_{-sp}^{+}M_{+sp}^{-}+M_{-sp}^{-}M_{+sp}^{+}]
\end{eqnarray}
If we set$$S^{\pm}=\pm \sin s +\sin(k_{x}-s)$$ and $$P^{\pm}=\mp
\sin p + \sin(k_{y}-p)$$
then the above Hamiltonian can be written
in the form
\begin{eqnarray}
H_{B}=\sum_{sp} 2iS^{+}(N_{+}^{+}M_{-}^{-}-M_{+}^{-}N_{-}^{+})
 +2iS^{-}(M_{+}^{+}N_{-}^{-}-N_{+}^{-}M_{-}^{+})\nonumber\\
-2P^{+}(M_{+}^{+}N_{-}^{+}-N_{+}^{+}M_{-}^{+})+2P^{-}
 (N_{+}^{-}M_{-}^{-}-M_{+}^{-}N_{-}^{-})\nonumber\\
-2m(M_{-}^{+}M_{+}^{-}+M_{-}^{-}M_{+}^{+}
 \end{eqnarray}
In matrix form we have:
\begin{equation}
\left(\begin{array}{cccc} 0 & -2m & -2P^{+} & -2iS^{-} \\-2m & 0
&2iS^{+} & 2P^{-} \\-2P^{+} & -2iS^{+} & 0 & 0 \\2iS^{-} & 2P^{-}
& 0 & 0
\end{array}\right)
\end{equation}
  From the above matrix the  the four eigenvalues $\lambda$ can easily be obtained as:
  \bea
 \lambda= \mp\sqrt{m^{2}+ \sin^{2}p + \sin^{2}s}\pm \sqrt{m^{2}+ \sin^{2}(k_{x}-s) +
  \sin^{2}(k_{y}-p)}
  \eea
  Note that as in the previous case of PA , equation (54) is
  consistent with the well-known solutions of the Dirac
  Hamiltonian on the lattice for single fermions.The procedure
  above generalizes to arbitrary large dimensions.
\section{Discussions}
In the SSH case, eqn.(9), the lattice-fermion interactions along
the chain led to bond-bosons that are irreducible of translations
by two sites. The bond-boson eigenenergies, however ,are exact
sums of energies of the component fermions.The mapping from
fermions to bosons is energetically neutral.
\newline
So far we have discussed the mapping of fermions to bond-bosons
for the hopping and for parts of fermion-lattice interactions.The
fermion-fermion interactions,such as the coulomb ;
\begin{equation}
H_{c}=\frac{1}{2}\sum
\alpha_{nm}(c_{n}^{\dag}c_{n})(c_{m}^{\dag}c_{m})
\end{equation}
needs to be written in terms of the bond bosons.Rewrite it as:
\begin{equation}
H_{c}=- \frac{1}{2}\sum
\alpha_{nm}(c_{n}^{\dag}c_{m}^{\dag})(c_{n}c_{m})
\end{equation}
Now, since
\begin{equation}
e_{\pm lk}=\sum_{p} c_{p}^{\dag}c_{p+l}^{\dag}e^{\pm ipk}
\end{equation}
Therefore , \bea c_{p}^{\dag}c_{p+l}^{\dag}=\sum_{k}
e_{+lk}e^{-ipk} \eea
 .Hence all bilinears of the type
$c_{m}^{\dag}c_{n}^{\dag}$ or $c_{m}c_{n}$ may be expressed in
terms of the bosonic translational eigenstates $e_{\pm lk}$. Hence
the interaction $H_{c}$ is expressible in terms of the
bond-bosons.Aside from the coulomb ,many other interactions may be
mapped to bond-bosons by the above method. The mapping yields a
quadratic in boson-bonds in place of the quartic fermions.
Similarly using the previous relations we can convert an
interaction term of the type $\sum
c_{i}^{\dag}c_{i}(b_{i}+b_{i}^{\dag})$, where $b_{i}$
are quantized lattice vibrations, into a form that has
interactions between $b_{i}$  with bond-bosons.The fermion gauge-boson
interactions $\sum c_{n}^{\dagger}c_{m}e^{i.\int_{n}^{m}{A.dl}}$ can be
 reduced in the same manner to interactions between the gauge-bosons and
the bond-bosons.
\newline
It is known from previous work[3] that the enveloping algebra for
fermions for the number-conserving case is $A_{n}$.It is also
known that the fermions belong to the fundamental (totally
antisymmetric) representations of $A_{n}$ For the lattice the
number of fermion states are determined by the number of lattice
points.The filling-factor determines, for this case , the specific
fundamental representation to which the fermions belong. The
lattice determines,to a large extent,the ways to combine the
generators of the enveloping algebra for it to be useful.For
instance the Cartan generators have to be suitably combined as
follows:the "distorted" generators $h_{ik}$ [1] do not commute
with each other.These have to be combined to construct hermitian
operators that constitute the Cartan subalgebra with the right
space-group properties to be useful on the lattice. Denote the
enveloping algebra on the lattice for the number conserving case
by Y.It has $A_{n}$ extended to include the discrete space
invariance.The induced representations are relevant.The physically
interesting boson mapping is constrained by the structure:
\newline
 $Y/A_{N}$=Discrete space symmetry
 \newpage
In case the underlying dynamics does not conserve the number,it is
known that the envelope algebra is $D_{n}$ ; the fermions belong
to the two irreducible spinor representations.The fundamental
representations of $A_{n}$ embed in these two spinor
representations of $D_{n}$. Once again the lattice extends the
algebra beyond $D_{n}$ to X  in order to include the discrete
space invariance, with the constraint that $D_{n}$ is an invariant
subalgebra of X and that
 $X/D_{n}$=the discrete space symmetry
\newline
The generators have to be suitably combined to identify physically
interesting boson degrees.These depend on the underlying space
group.The physically interesting boson-bond configurations for the
SSH-case was to illustrate some of these ideas.
\newline
As we go beyond 1+1 dimension into higher dimensions these ideas
are generalized. Once again, the underlying lattice space
invariance dictates how to combine the elements of the algebra.The
illustration in section 3 is but one such.In general the useful
combinations of generators that are bosonic depend on the space
symmetry. The induced representations are the ones that are
important.
\section{Acknowledgements}
One of us,J.C., acknowledges useful communications from Susumu
Okubo , Robert Gilmore and Eugene Marshalek on the subject.

\end{document}